\documentclass[proceedings]{JHEP3}

\PrHEP{PrHEP hep2001}                   
\conference{International Europhysics Conference on HEP}                

\usepackage{epsfig}                   
\def\s[#1,#2]{[#1\stackrel{\star}{,}#2]}         
\def\eq#1\en{\begin{equation}#1\end{equation}}

\def\sDD{{\,\,/\!\!\!\!{D}}}
\def\pp#1{\partial_#1}
\def\DD{\mathcal{D}}
\def\AA{\mathcal{A}}

\title{Non-Abelian gauge theory on noncommutative spaces}

\author{Peter Schupp  
        \\ 
        Sektion Physik, Universit\"at M\"unchen, Theresienstr.\ 37, 80333 M\"unchen,
	Germany\\                                          
        E-mail: \email{schupp@theorie.physik.uni-muenchen.de}}

\abstract{We present a brief introduction to the construction of gauge
theories on noncommutative spaces with star products. 
Particular emphasis is given to issues related to non-Abelian gauge groups 
and charge quantization. 
This talk is based on joined work with B. Jur\v co, J. Madore, L. M\"oller,
S. Schraml and J. Wess. }

\begin{document}

\section{Introduction}

The topic of this talk is the type of noncommutative gauge theory that
has become a recent focus of interest in string theory, where it appears
when one considers open strings in the presence of a background 
$B$-field~\cite{Seiberg:1999vs}.
In the following we are primarily interested in these  noncommutative
gauge theories as field theories, so it seems appropriate to present
an intuitive approach to their construction that is independent of 
string theory (but closely related to ideas of 
matrix theory)~\cite{Madore:2000en,Bichl:2001gu,Jurco:2001rq}.

A characteristic feature of noncommutative gauge theories is the
emergence of many new interactions. This includes self-couplings
of gauge bosons and may even include couplings between the photon
and neutral particles. One may picture these new interactions as
arising from the interplay of the gauge fields and noncommutative
space-time. The formalism that we shall present is particularily 
well-suited to capture this phenomenon. It is also the only known
approach that works for arbitrary gauge groups and representations. 

\section{Gauge theory on  noncommutative spaces}

When trying to replace the notions and concepts of commutative geometry in
the more general noncommutative framework the basic strategy is to not 
consider the space-time manifold itself but rather the algebra of 
functions on it.
In the noncommutative realm this algebra  is replaced by an 
arbitrary associative algebra. We shall refer to elements of this algebra,
e.g., fields $\widehat\Psi$, $\hat A_\mu$, $\hat F_{\mu\nu}$, 
gauge parameters $\hat\Lambda$
and coordinates $x^\mu$,
as ``functions on noncommutative space-time.'' 
The simplest example of such a noncommutative space is given
by the canonical structure
$\s[x^\mu,x^\nu] = i \theta^{\mu\nu}$,
with a constant antisymmetric matrix~$\theta^{\mu\nu}$.  One could of
course have more complicated structures, e.g., with commutation relations
that close linearily (Lie structure) or quadratically (quantum space
structure). 
A priori there is no reason
to  expect that~$\theta^{\mu\nu}$
is constant, but we shall concentrate on that 
case in the following for simplicity of presentation.
We use the symbol~$\star$ to denote the product of
the noncommutative structure; this does not need to be
a star product, but we are here especially interested 
in noncommutative structures that have a well-defined classical
limit and lend themselves to a perturbative formulation,
as is the case for star products.\\[1em]
The construction of a gauge theory on a given 
non-commutative space can be based 
on a few fundamental ideas: the concept of covariant coordinates, 
the requirement of locality, gauge equivalence 
and consistency conditions. 

\subsection{Covariant coordinates}

The infinitesimal 
non-commutative gauge 
transformation
of a fundamental matter field is 
\eq
\hat\delta\widehat\Psi = i \hat\Lambda \star \widehat\Psi .
\label{deltapsi}
\en
In the non-Abelian case the symbol $\star$ includes also matrix multiplication.
We observe that multiplying $\widehat\Psi$ on the left by a coordinate $x^\mu$ 
is not a covariant operation because the gauge 
parameter does not commute with it:
\[
\hat\delta(x^\mu\star\widehat\Psi)= ix^\mu\star\hat\Lambda\star\widehat\Psi
\neq i\hat\Lambda\star x^\mu\star\widehat\Psi.
\]
In complete analogy to the covariant 
derivatives of ordinary gauge theory we need to introduce
covariant coordinates $X^\mu = x^\mu + \hat A^\mu$ 
where $\hat A^\mu$ is a 
non-commutative analog of the gauge potential.
In the case of constant, non-degenerate $\theta^{\mu\nu}$ it is more convenient
to work with $\hat A_\nu$, where  
$\hat A^\mu=\theta^{\mu\nu} \hat A_\nu$, with
\eq
\hat\delta \hat A_\mu = \pp\mu\hat\Lambda + i\s[\hat\Lambda,\hat A_\mu].
\label{deltaA}
\en
Similarily, covariant 
functions $\DD(f) = f + \AA(f)$ can be introduced~\cite{Jurco:2000fb}. 
From the covariant coordinates one can construct further covariant objects including the
non-commuatative field strenght
\eq
\widehat F_{\mu\nu}  = \pp\mu\hat A_\nu - \pp\nu\hat A_\mu  -i\s[\hat A_\mu,\hat A_\nu],
\qquad \hat\delta\widehat F_{\mu\nu} = i\s[\hat\Lambda,\widehat F_{\mu\nu}],
\en
related to the commutator of covariant coordinates, and the covariant
derivative
\eq
\widehat D_\mu \widehat\Psi = \pp\mu \widehat\Psi + i \hat A_\mu\star\widehat\Psi,
\label{covder}
\en
related to the covariant expression $X^\mu \star \widehat\Psi - \widehat\Psi\star x^\mu$.

\subsection{Classical limit and locality} 

A star product of two functions $f$, $g$
is a power series in a formal order parameter $h$
starting with the commutative product plus
higher order terms chosen in such a way as to yield an associative product.
It can be seen as a tower build upon its leading term 
that is determined by a Poisson tensor $\theta^{\mu\nu}$:  
\eq
f \star g = 
f \cdot g + \frac{i h}{2}\theta^{\mu\nu} \pp\mu g \cdot \pp\nu f +\ldots
\en
It is a natural to  ask whever it is possible to express also the
non-commutative fields $\hat A$, $\widehat \Psi$ and non-commutative 
gauge parameter $\hat\Lambda$
in a similar fashion
as towers build upon the corresponding ordinary fields $A$, $\Psi$  
and ordinary
gauge parameter $\Lambda$. This is indeed the case; there 
are so-called Seiberg-Witten
maps \cite{Seiberg:1999vs} that express the non-commutative 
quantities as local 
functions of the ordinary fields:
\begin{eqnarray}
\hat A_\mu & = & A_\mu + \frac{1}{2} \theta^{\xi\nu}
(A_\nu\pp\xi A_\mu + 
F_{\xi\mu}A_\nu) + \ldots  \label{SWA} \\ 
\widehat \Psi & = & \Psi + \frac{1}{2} \theta^{\mu\nu}A_\nu\pp\mu\Psi
 + \ldots \label{SWPsi} \\
\hat\Lambda & = & \Lambda 
+ \frac{1}{2} \theta^{\mu\nu} A_\nu\pp\mu \Lambda+ \ldots
\label{SWLambda}
\end{eqnarray}
where $F_{\mu\nu} = \pp\mu A_\nu - \pp\nu A_\mu -i [A_\mu,A_\nu]$ is the
ordinary field strength.
By a  local
function of a field we mean a formal series in the 
deformation parameter $h$ that at each order in  $h$ depends on the field
and a finite number of derivatives of the field.
We shall henceforth use a hat $\,\widehat\;\,$ to denote non-commutative
quantities that are supposed to be expanded as local functions of 
their classical counterparts via Seiberg-Witten maps.

\subsection{Gauge equivalence and consistency condition}

The Seiberg-Witten maps (\ref{SWA})--(\ref{SWLambda})
have the remarkable property that ordinary gauge
transformations $\delta_\Lambda A_\mu = \pp\mu \Lambda + i[\Lambda,A_\mu]$
and $\delta_\Lambda \Psi = i \Lambda\cdot \Psi$ induce 
noncommutative gauge transformations
(\ref{deltapsi}), (\ref{deltaA}) of 
$\hat A$, $\widehat \Psi$, $\hat \Lambda$. 
Furthermore, any pair of  non-commutative gauge parameters 
$\hat\Lambda$, $\widehat\Sigma$ has to satisfies the following consistency condition
\cite{Jurco:2001rq}
\eq
\s[\hat\Lambda,\widehat\Sigma] + i \delta_\Lambda \widehat\Sigma
- i \delta_\Sigma \hat\Lambda = \widehat{[\Lambda,\Sigma]}.
\label{consistent}
\en
The gauge equivalence and consistency conditions do not uniquely determine
Seiberg-Witten maps. To the order considered here we have the
freedom of classical field redefinitions and noncommutative 
gauge transformations.
We have used that freedom to obtain the particularly simple set of
maps. Other choices may be
more convenient in applications. A version with hermitean
$\hat\Lambda$ can, e.g., be obtained from  (\ref{SWA})--(\ref{SWLambda}) 
by a noncommutative gauge transformation generated
by $\frac{1}{4}\theta^{\mu\nu} a_\mu a_\nu$. 

The freedom in the Seiberg-Witten map
is essential for the renormalization of noncommutative gauge
theory~\cite{Bichl:2001cq}.
It is also important in the context of tensor products of gauge groups.  
For instance, for a field $\Phi$ that transforms on the left and on
the right under two arbitrary gauge groups,
we have the following hybrid Seiberg-Witten map,
\eq
\widehat\Phi = \Phi + \frac{1}{2} \theta^{\mu\nu}a_\nu\pp\mu\Phi
+ \frac{1}{2} \theta^{\mu\nu}\pp\mu\Phi\tilde a_\nu + 
\frac{i}{2} \theta^{\mu\nu} a_\nu \Phi \widetilde a_\nu + \ldots.
\en
Under
$\delta\Phi = i\lambda\Phi + \Phi(i\tilde\lambda)^\dagger$,
$\delta a_\nu = \pp\nu\lambda + i[\lambda,a_\nu]$,
$\delta \tilde a_\nu = \pp\nu\tilde \lambda + i[\tilde \lambda,\tilde a_\nu]$
we find 
\eq
\delta \widehat\Phi = i \hat\lambda \star \widehat\Phi
+ \widehat\Phi\star (i \widehat{\tilde\lambda})^\dagger.
\en

\section{Non-Abelian gauge groups in noncommutative setting}

The commutator of two Lie algebra-valued noncommutative gauge parameters,
\eq
\s[\hat\Lambda,\hat\Lambda'] = \frac{1}{2}\{\Lambda_a(x)\stackrel{\star}{,}
\Lambda'_b(x)\}[T^a,T^b] + \frac{1}{2}\s[\Lambda_a(x),\Lambda'_b(x)]\{T^a,T^b\},
\label{com}
\en
does not close in the Lie algebra, because 
the coefficient  of $\{T^a,T^b\}$
is in general non-zero. (The only important exception is
$U(N)$ in the fundamental representation.) 
We thus have to consider enveloping algebra-valued noncommutative gauge  parameters
\eq
\hat\Lambda = \Lambda^0_a(x) T^a +  \Lambda^1_{ab}(x) :T^a T^b:
+ \Lambda^2_{abc}(x) :T^a T^b T^c: + \ldots,
\en
 and fields $\hat A_\mu$ \cite{Madore:2000en}.
A priori, it appears
that we then have an infinite number of degrees of freedom.
Via the Seiberg-Witten map, however, all the terms in
$\hat\Lambda$ and $\hat A_\mu$ can
be expressed in terms of a finite number of classical
parameters and fields.

\subsection*{Noncommutative Yang-Mills action}

For constant $\theta$ the ordinary integral is a trace for the $\star$-product:
$\int f \star g  = \int g \star f = \int f g$.
An invariant action for the gauge potential and the matter fields is
$$
S =  \int d^4x\left[-\frac{1}{2g^2}\mbox{tr} \widehat F_{\mu\nu} \star 
\widehat F^{\mu\nu} 
+ \overline{\widehat\Psi}
\star (i\gamma^\mu \widehat D_\mu - m) \widehat\Psi\right], 
$$
where  
$\widehat D_\mu \widehat\Psi \equiv \pp \mu \widehat\Psi - i \hat A_\mu \star
\widehat\Psi$, $\widehat F_{\mu\nu} = \pp \mu \hat A_\nu - \pp \nu \hat A_\mu - i\s[\hat A_\mu,\hat A_\nu]$.
Expanding $\hat A_\mu$ and $\widehat\Psi$ to first order in $\theta$ using a hermitean version of
the Seiberg-Witten map yields 
\begin{eqnarray*}
S & =  & \int d^4x\left[-\frac{1}{2g^2}\mbox{tr} 
F_{\mu\nu}  F^{\mu\nu} +\frac{1}{4g^2} \theta^{\mu\nu} \mbox{tr} F_{\mu\nu}
F_{\rho\sigma}F^{\rho\sigma} - \frac{1}{g^2} \theta^{\mu\nu} \mbox{tr} F_{\mu\rho} F_{\nu\sigma}
F^{\nu\sigma} \right.\nonumber\\
&&+ \left.\overline{\psi}(i\sDD -m)\psi -\frac{1}{4} \theta^{\mu\nu}
\overline{\psi} F_{\mu\nu}(i\sDD -m)\psi -\frac{i}{2}
\theta^{\mu\nu} \overline{\psi} \gamma^\rho F_{\rho\mu}  D_\nu \psi
\phantom{\textstyle{\frac{1}{g^2}}}\!\!\!\right]
\end{eqnarray*}
with $D_\mu \psi \equiv \pp \mu \psi - i A_\mu \psi$ and $F_{\mu\nu} = \pp \mu A_\nu - \pp \nu
A_\mu - i[A_\mu,A_\nu]$.

\section{Charge in noncommutative QED}

The only couplings in addition to (\ref{covder})
of the noncommutative gauge boson $\hat A_\mu$
to a matter field $\widehat \Psi$ compatible with the non-commutative
gauge transformation (\ref{deltaA}) 
are
$\pp\mu \widehat\Psi - i\widehat\Psi\star\hat A_\mu$
and $\pp\mu \widehat\Psi +i\s[\hat A_\mu,\widehat\Psi]$.
It thus appears that only U(1) charges $1$, $-1$, $0$
are possible.
(The latter possibility shows
how a neutral particle can couple to an abelian gauge field in 
a noncommutative setting.)
We should of course consider physical fields $\hat a^{(n)}_\mu(x)$. 
Let $Q$ be the generator of $U(1)$ (charge operator), $e$ 
a coupling constant and $\psi^{(n)}$ a field 
for a particle of charge $q^{(n)}$.
Then 
$A_\mu  = e Q a_\mu(x)$ and
$\hat A_\mu \star\hat\psi^{(n)} = e q^{(n)}\hat a^{(n)}_\mu(x)\star\hat\psi^{(n)}$,
since the Seiberg-Witten map $\hat A_\mu$ depends explictly on $Q$.
In ordinary QED there is only one photon, i.e., there is 
no need for a label $(n)$ on $a_\mu$.
Here, however, we have a separate $\hat a^{(n)}_\mu$ for every charge
$q^{(n)}$ in the theory, because
due to the $\star$-commutator in the transformation of $\hat a_\mu^{(n)}$
it is not possible to absorb $q^{(n)}$ in a redefinition of $\hat a_\mu^{(n)}$.
We can have any charge now, but
it appears that we have
too many degrees of freedom. This is not the case, however, since all
$\hat a^{(n)}_\mu$ are local functions of the correct number of
classical gauge fields $a_\mu$ via the Seiberg-Witten map (\ref{SWA}).  

\section{Construction of the Seiberg-Witten map}

In the Abelian case the Seiberg-Witten map is known explicitly for any
Poisson structure $\theta(x)$ and corresponding Kontsevich $\star$-product
\cite{Jurco:2000fb}.
The construction is based on equivalent star products $\star$, $\star'$
that are quantizations of Poisson structures $\theta$ and $\theta'
= \theta(1+F\theta)^{-1}$. 
There is also a path-integral formulation based on noncommutative
Wilson lines \cite{Okuyama:2000ig}. 
The non-Abelian case is technically
more involved and only formally related to the Abelian
case. For a cohomological approach based
on the consistency relation (\ref{consistent}) see \cite{Brace:2001rd}.

\section{Finite gauge transformations and noncommutative vector bundle}

The infinitesimal gauge parameter 
$\hat\lambda \equiv \hat\lambda_{[a]}$ can be promoted to
a full finite noncommutative gauge transformation
$\widehat g_{[a]} = \exp_\star({\delta_\lambda})
\star \exp_\star({-\delta_\lambda + i\hat\lambda_{[a]}})$ 
corresponding to a group element $g = e^{i\lambda}$. The consistency
relation (\ref{consistent}) now becomes a ``noncommutative group law''
\cite{Jurco:2001kp}
\eq
\widehat{g_1}_{[a_{g_2}]} \star \widehat{g_2}_{[a]} 
= \widehat{g_1\cdot g_2}_{[a]},
\en
with the gauge transformed gauge potential $a_{g_2}$ in the first factor. 
The $\hat g_{[a]}$  can be used as transition functions in the construction 
of noncommutative vector bundles, which
are the underlying mathematical structure of the noncommutative
gauge theories that we have been considering \cite{Jurco:2001kp}.

\subsection*{Acknowledgements}
I would like to thank B. Jur\v co, J. Madore, L. M\"oller,
S. Schraml and J. Wess for fruitfull collaboration and
P. Aschieri and W. Behr for helpful discussions.

\end{document}